# Wireless Electronic-free Mechanical Metamaterial Implants

Jianzhe Luo[1,2], Wenyun Lu[1,2], Pengcheng Jiao[3], Daeik Jang[2], Kaveh Barri[4], Jiajun Wang[3], Wenxuan Meng[2], Rohit Prem Kumar[5], Nitin Agarwal[5,6,7], D. Kojo Hamilton[5,6,7], Zhong Lin Wang[8,9,10], Amir H. Alavi[1,2,a]

**Abstract**

Wireless force sensing in smart implants enables real-time monitoring of mechanical forces and facilitates dynamic adjustments to optimize implant functionality in-situ. This capability enhances the precision of diagnostics and treatment leading to superior surgical outcomes. Despite significant advancements in wireless smart implants over the last two decades, current implantable devices still operate passively and require additional electronic modules for wireless transmission of the stored biological data. To address these challenges, we propose an innovative wireless force sensing paradigm for implantable systems through the integration of mechanical metamaterials and nano energy harvesting technologies. We demonstrate composite mechanical metamaterial implants capable of serving as all-in-one wireless force sensing units, incorporating functions for power generation, sensing and transmission with ultra-low power requirements. In this alternative communication approach, the electrical signals harvested by the implants from mechanical stimuli are utilized directly for the wireless transmission of the sensed data. We conduct experimental and theoretical studies to demonstrate the wireless detection of the generated strain-induced polarization electric field using electrodes. The feasibility of the proposed wireless force sensing approach is evaluated through a proof-of-concept orthopedic implant in the form of a total knee replacement. The findings indicate that the created wireless, electronic-free metamaterial implants with a power output as low as 0.1 picowatts enable direct, self-powered wireless communication during force sensing across air, simulated body fluid and animal tissue. We validate the functionality of the proposed implants through a series of experiments conducted on an *ex vivo* human cadaver knee specimen. Furthermore, the effect of electrode size and placement on the strength of the received signals is examined. Finally, we highlight the potential of our approach to create a diverse array of mechanically-tunable implants capable of precise force measurements and wireless real-time data transmission, all without relying on any external antennas, power sources, or telemetry systems.

## Introduction

Smart wireless implants have emerged as a progressive advancement in modern healthcare. They enable continuous monitoring of a spectrum of physiological signals [1,2]. These implantable devices possess the potential to fundamentally transform patient care via enabling the continuous acquisition of data and facilitating timely medical interventions [3,4]. Meanwhile, force sensing plays a pivotal role in the realm of smart implants. This process encompasses applications such as intraocular pressure monitoring, assessment of joint biomechanics, and stabilization of orthopedic implants [5-7]. However, the conventional methods for acquiring force sensing data from smart implants (e.g., LC resonant [8], magnetic soft material [3], electromagnetic (EM) waves [1]) require various bulky modules for signal generation, power supply, signal modulation, and transmission [1]. The utilization of external electronics or batteries in biomedical implants proves impractical due to limitations in their operational lifespan, size, and associated chemical risks [9]. Over the past four decades, substantial research has been conducted to advance the development of smart implants with force sensing capabilities, particularly in the field of orthopedics

[1]*Department of Bioengineering, University of Pittsburgh, Pittsburgh, PA, USA*
[2]*Department of Civil and Environmental Engineering, University of Pittsburgh, Pittsburgh, PA, USA*
[3]*Ocean College, Zhejiang University, Zhoushan, Zhejiang, China*
[4]*Department of Civil and Systems Engineering, Johns Hopkins University, Baltimore, MD, USA*
[5]*Department of Neurological Surgery, University of Pittsburgh School of Medicine, Pittsburgh, PA, USA*
[6]*Department of Neurological Surgery, University of Pittsburgh Medical Center, Pittsburgh, PA, USA*
[7]*Neurological Surgery, Veterans Affairs Pittsburgh Healthcare System, Pittsburgh, PA, USA*
[8]*Beijing Institute of Nanoenergy and Nanosystems, Chinese Academy of Sciences, Beijing, China*
[9]*School of Materials Science and Engineering, Georgia Institute of Technology, Atlanta, GA, USA*
[10]*Yonsei Frontier Lab, Yonsei University, Seoul 03722, Republic of Korea*
[a]Email: alavi@pitt.edu (A.H. Alavi)

[10]. Fig. 4a shows the evolution of force sensing smart implants, highlighting key studies conducted in this field [5-7,9,11-57]. Various types of smart implants investigated are wired [17,18], battery-powered wireless [5,19-21], passively-powered wireless [22-57], and self-powered wireless equipped with microcontrollers [6,7,9,11-16]. The estimated operational power of these existing implants, including transmission needs, ranges from 1 µW to 10 mW [58]. Smart implants incorporating passively-powered wireless capabilities have demonstrated the most enduring success. Most of these passive implants utilize radio-frequency identification (RFID) technology for wireless power supply and sensor interrogation. However, the RFID approach faces considerable limitations in tissue environments [9,59,60]. In addition, these passive wireless implants are incapable of continuously recording the data unless exposed to an inductive energy source. They are often designed to capture momentary changes in force/strain levels, offering only a single snapshot in time [3,9]. The majority of studies concentrating on this class of implants were conducted between 2000 and 2015 (Fig. 4a). Recent studies have explored the concept of harvesting energy from human motion to create self-powered implants. The latest generation of these self-powered implants, such as piezo-floating-gate (PFG) [11-13] and Fowler-Nordheim (FN) [9,61] implants, offer partial solutions to the challenges of passive implants. However, they remain electronic modules that require additional RFID or ultrasound connectivity to transmit the stored data. Methods such as galvanic coupling [62] and ionic communication [63] have emerged as promising approaches for wireless intrabody data transmission. These techniques typically employ pairs of electrodes, with one acting as the signal transmitter and the other as the receiver, to establish a communication link across tissue [62-64]. These technologies have not yet been applied to wireless force sensing. Similar to other wireless force sensing technologies, they rely on external power sources and onboard microcontrollers to facilitate communication between the transmitter and receiver electrodes. Consequently, there is a growing demand for wireless force sensing techniques implants characterized by their compact form, self-powered operation, and autonomous data transmission.

In recent decades, significant research efforts have focused on developing smart biomaterials that can mimic the properties of human tissues [65]. Initially, the focus was primarily on enhancing the mechanical performance of these biomaterials. Subsequently, the concept of mechanical metamaterials, artificial structures endowed with specific properties not encountered in nature, was introduced to augment mechanical, physical, and biological characteristics [6,7,66-68]. For instance, Zadpoor et al. [68] highlighted the potential of their proposed mechanical metamaterials in tissue replacement, thereby facilitating tissue regeneration. In our previous study [7], we introduced multifunctional metamaterial implantable devices capable of sensing spinal forces, harvesting energy from spinal motion, and monitoring bone healing progress. However, a significant research gap remains regarding the establishment of a wireless communication paradigm for retrieving the biological data collected by such systems. This advancement would enhance the suitability of these materials for various biomedical applications.

Here, we introduce a new concept based on Maxwell's displacement current to realize wireless communication directly using mechanical metamaterial implants. We develop proof-of-concept metamaterial orthopedic implants that can harvest energy from body motions and use the generated electrical signal for "direct, wireless and electronic-free" transmission of the sensed data, without relying on additional electronics. These implants enable mid-range wireless communication in real-time with power outputs in the picowatt (pW) range. Experimental studies are conducted under various loading conditions to evaluate the communication capabilities of these all-in-one wireless metamaterial implants. We present theoretical models to characterize the strain-induced polarization electric field generated by these metamaterial implants across different media. The functionality of the proposed implants is further studied through a series of experiments conducted using a human cadaver knee specimen. Finally, we discuss the future of personalized electronic-free wireless metamaterial implants capable of accurately measuring the forces and wirelessly transmitting real-time data.

**Results**

We present an innovative signal transmission mechanism for wireless force sensing. Our approach transforms implants into entirely self-contained units capable of wirelessly transmitting the senses data with ultra-low power requirements, operating in the pW range (Fig. 4a). This is achieved by integrating mechanical metamaterials with nano energy harvesting technologies to create a composite biomaterial. These composite material systems are constructed from a combination of conductive and dielectric lattices, specifically designed to induce triboelectrification. This allows them to function as triboelectric nanogenerators (TENGs) [69-73] when subjected to applied forces. While our previous work explored the versatility of this metamaterial platform for creating scalable structural systems with sensing capabilities [6,7], this research tackles a distinct, longstanding challenge in the biomedical field: *achieving direct wireless, electronic-free interrogation of implants*. The central question we aim to answer is "*can composite mechanical metamaterials enable the wireless transmission of self-generated electrical signals without the need for integrated electronics and external power sources?*". The vision for this research is shown in Figs. 1b-d. We aim to understand the underlying mechanisms and fundamental principles necessary for the development of self-powered electronic-free wireless metamaterial implants. The self-powering wireless capability potentially enables miniaturization of implants by eliminating the requirement for external power sources, extra electronics, or large antennas. Fig. 1b shows the schematics of conductive and dielectric lattices forming a wireless mechanical metamaterial lattice with wireless communication functionality. Upon mechanical triggering, contact-electrification occurs within the metamaterial lattice, leading to the generation of an electric signal proportional to the applied force. The strain-induced signal produced by the lattice can then be detected by an electrode wirelessly (Fig. 1b).

Characterizing wireless transmission of the strain-induced signals generated by the wireless metamaterial systems is a challenging task. Our hypothesis revolves around the utilization of Maxwell's displacement current to formulate wireless transmission of the measured force signals. This concept can be elucidated using a capacitance model [74,75], as shown in Fig. 1c. The analysis of electric field propagation is approached by considering the displacement current, akin to the principles governing TENG [74-77]. In this model, the transmitting and receiving electrodes serve as the positive and negative terminals of a capacitor, while the intervening medium acts as the dielectric. Under the influence of the electric field (E), the dielectric becomes polarized, generating a polarization electric field (P). This polarization electric field arises from the juxtaposition of negative and positive polarization charges [75]. The resulting combined electric field (E′) can be quantified relative to E by defining the relative permittivity ($\varepsilon_r$) [75]:

$$\varepsilon_r = E/E' \tag{1}$$

The relationship between polarization charge (Q′) and the charge (Q) on the transmitting electrode can be defined as:

$$Q' = (1 - 1/\varepsilon_r)Q \tag{2}$$

Owing to the attenuation of the electric field during propagation through the medium, the charge received (Q″) at the receiving electrode is less than Q′. The Gauss's law of Maxwell's equations gives the relationship between the electric displacement vector D and distribution of free charges in space $\rho$ as [75]:

$$\nabla \cdot \boldsymbol{D} = \rho. \tag{3}$$

$\boldsymbol{D}$ can be expressed as:

$$\boldsymbol{D} = \varepsilon_0 \boldsymbol{E} + \boldsymbol{P} \tag{4}$$

where D signifies the electric displacement vector, $\varepsilon_0$ is the permittivity in a vacuum, and $\boldsymbol{P}$ represents the medium polarization vector, respectively [75]. In practical scenarios, polarization can also result from the strain field, which emerges due to surface contact-electrification (e.g., triboelectric effect) and is independent of the presence of an electric field [75-77]. To incorporate the influence of contact-electrification-induced electrostatic charges into Maxwell's equations, Wang [75] introduced an additional term ($\boldsymbol{P}_s$) that represents the polarization arising from the relative movement of the charged dielectric media, the final displacement vector is given in Eq. (5):

$$\boldsymbol{D} = \varepsilon_0 \boldsymbol{E} + \boldsymbol{P} + \boldsymbol{P}_s \tag{5}$$

In Eq. (5), polarization vector $\boldsymbol{P}$ arises from the impact of an external electric field, while $\boldsymbol{P}_s$ predominantly emerges from the presence of surface charges that are independent of the electric field's existence [75]. The displacement current observed in TENG is given by $\frac{\partial}{\partial t}\boldsymbol{P}_s$. Based on these explanations, we hypothesize that the strain-induced polarization term, i.e. $\boldsymbol{P}_s$, can be employed to develop a current transport equation for wireless metamaterial systems, thereby enabling wireless transmission functionality.

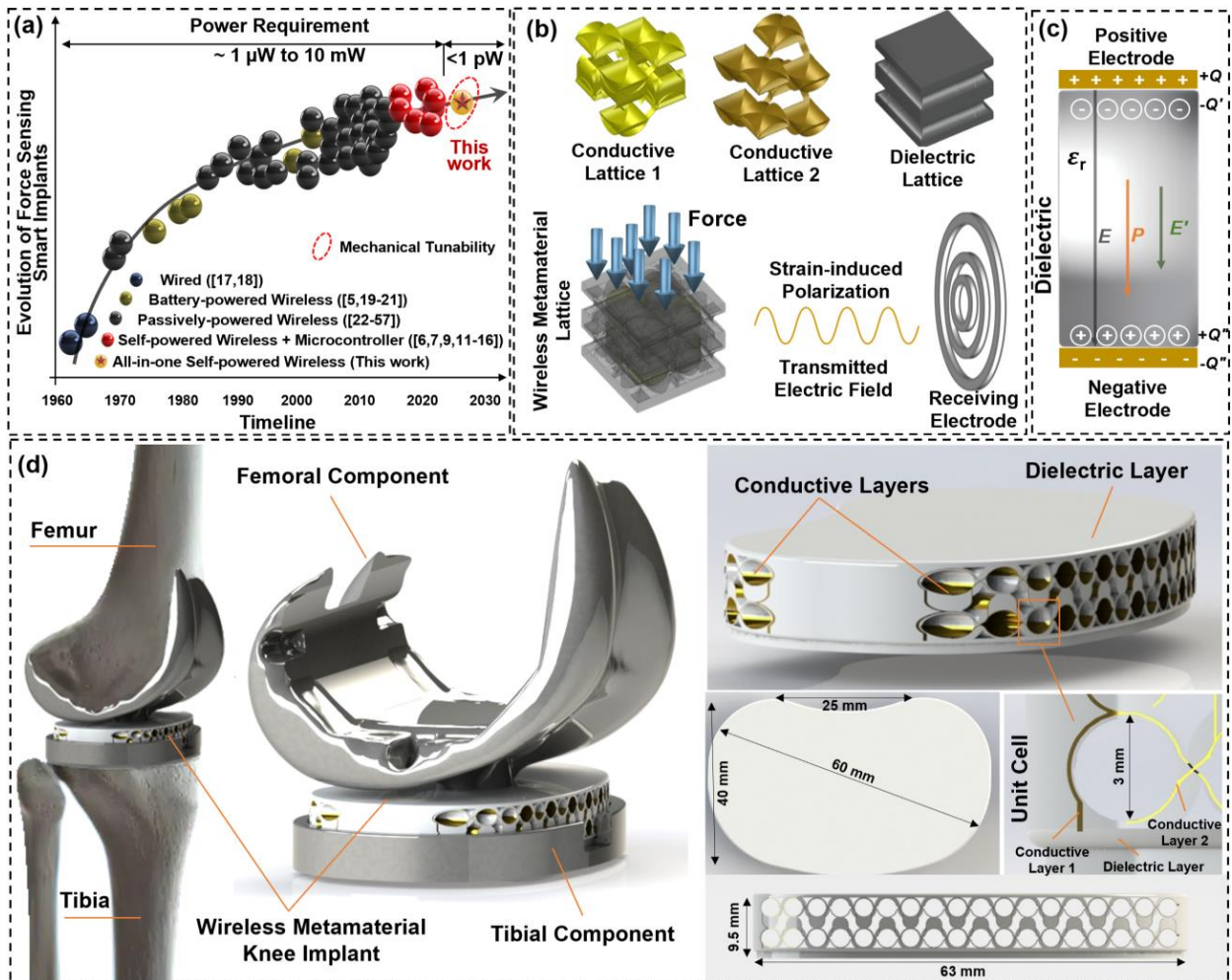

**Fig. 1. Developing self-powered, electronic-free metamaterial implants for wireless force sensing.** (a) Evolution of smart implants equipped with force sensing, demonstrating key studies conducted in this field, functionalities of the implants and their power demand. The proposed implants enable mid-range wireless communication in real-time with power outputs in the pW range. They are also the only class of mechanically tunable implants, a unique feature attributed to their metamaterial nature. (b) Schematics of a wireless metamaterial system composed of rationally designed conductive and dielectric lattices. This multi-material composite metamaterial induces contact-electrification under mechanical triggering. The generated signal is proportional to the applied force. The strain–induced Maxwell's displacement current generated by the wireless metamaterial lattice enables wireless transmission of the sensed signal through various media without any antenna and power supply. (c) Working principle for wireless communication with wireless metamaterial lattice. It operates as a capacitance model, as reported in [76]. (d) Schematics and dimensions of a mechanically-tunable wireless metamaterial TKR implant capable of self-powering through knee loads, measuring forces, and wirelessly transmitting real-time data, without relying on any battery-powered telemetry system or external antennas.

We demonstrate the viability of the proposed approach using a proof-of-concept total knee replacement (TKR) implant with wireless force sensing. Fig. 1d illustrates the schematics and dimensions of the wireless metamaterial TKR implant. We chose TKR implants since they are exposed to significant loading during daily activities like jumping, running, and walking [78]. The TKR surgery numbers are rapidly growing and expected to reach 3.48 million by 2030 in the United States [79]. This surgery is performed to remove damaged cartilage and bone, and to replace the removed part with artificial components (e.g., tibial, femoral or spacers), known as knee implants [80,81]. The knee implants generate a new surface between tibia and femur. Approximately, 20% of the patients with TKR surgery experience pain and reduced functionality levels after the surgery [82]. A main contributor to the unsatisfied function of TKRs is incorrect ligament balancing, which can accelerate abrasion from unbalanced joint reaction force and increase prosthetic loosening [80,81]. Although various techniques are used for analyzing post-surgery TKRs kinematics, direct measurement of loads on the TKR implant components is still challenging. These loads can eventually result in implant failure. A better understanding of the loading pattern on knee implants can be clinically beneficial to assess the health and functionality of the prosthesis [83]. All of the current smart TKR implants with force-sensing capabilities contain multiple electronic modules and rely on inductive wireless power transfer [84]. Also, the only clinically available smart TKR implant, Persona IQ, launched in 2021 by Zimmer Biomet and Canary Medical [85,86], is a tibial stem instrumented with internal motion sensors and battery-powered telemetry modules to collect and transmit kinematic data. Despite the novelty of the Persona IQ implant, it requires a 58 mm long tibial stem extension to house electronics, resulting in additional bone resection and an alteration to the implant orientation [86].

A mechanically-tunable TKR implant with the ability to self-power through knee loads, accurately measure forces, and wirelessly transmit real-time data, all without relying on a battery-powered telemetry system, holds the potential for substantial clinical advantages. Herein, we design a wireless metamaterial TKR polymer spacer as it directly carries the knee load between the tibial and femoral components and thus can provide objective information about the loading pattern on the implant (Fig. 1d). The TKR implants are specifically shaped to mimic the natural anatomy of a healthy knee, which is essential for ensuring a precise fit, enhancing joint stability, and promoting even weight distribution across the joint surface [87,88]. Preliminary finite element (FE) simulations were carried out to design the multi-layered metamaterial spacer with an elastic modulus within the range of central region of human medial meniscus (~20-80 kPa) [89,90]. The wireless metamaterial TKR implant model was then 3D printed and tested to determine its mechanical and electrical properties. The fabrication details are provided in Materials and Methods. During the tests, the implant was placed within a polyacrylonitrile testing bath with a diameter of 15 cm. A conductive copper tape with a width of 5 cm was affixed to the exterior of the bath to serve as the receiving electrode. The printed implant and test setup are shown in Fig. 2a. The elastic modulus was calculated following the procedure explained in the ISO standard 13313:2011 for porous and cellular metals [91]. Fig. 2b shows the obtained stress–strain curves for the implant. Video S1 in Supplementary Materials shows the deformation of the implant under uniaxial loading. The slope of the fitted straight lines represents the elastic modulus value (~ 45 kPa). Depending on the clinical requirements, such metamaterial implants can be designed with any desired mechanical properties, as reported in [7].

The implant was initially tested under uniaxial loading conditions at 1, 3, and 5 Hz, corresponding to knee joint loading frequencies during walking and running [32,92]. Knee joint load ranges from 1.8 times body weight to 8.1 times body weight during daily life activities [92]. To assess the mechanical and electrical performance of the metamaterial spacer, cyclic loading tests were conducted within the range of 0 to 350 N (~ 4 times the body weight of a 100 kg person). Figs. 2c-e show the electrical signal transmitted by the implant lattice and the received by the electrode in air under uniaxial loading at different frequencies. Referring to Fig. 2c, the coefficient of determination ($R^2$) between the wirelessly received voltage signal and the applied force, increasing from 0 to 350 N, is high ($R^2$ = 0.99), indicating that the implant functions as a force sensor. Videos S2-4 in Supplementary Materials show typical signals transmitted and received under uniaxial loading at 1, 3 and 5 Hz, respectively. The typical applied loading cycles at 1 Hz are also illustrated in Fig. 2c. The electric signals are proportional to the applied force, as thoroughly discussed in our prior study [6,7]. Referring to Figs. 2c-e, is evident that both the transmitted voltage signals and their corresponding received signals increase with the rising frequency. Here, we define "signal delivery ratio" as the ratio of voltage signal received to the signal transmitted. The signal delivery ratios are 0.52, 0.44 and 0.36 at 1, 3 and 5 Hz, respectively. The implant was then tested under internal rotation and varus loading conditions at 1 Hz. A 25° internal rotation was taken into consideration, as suggested in [93]. The results are shown in Figs. 2f and g. The signal delivery ratios are 0.61 and 0.39 for the internal rotation and varus loading, respectively. The output performance of the wireless metamaterial TKR implant can be determined by measuring open-circuit voltage, current, and power density. The power-voltage-current curve output of the implant is shown in Fig. 2h. The power output of the implant gradually increases from 0.1 MΩ to 100 MΩ, reaching a maximum value of 0.125 pW. The meta-mechatronic systems feature a built-in TENG mechanism, thereby providing high voltage and low current. This characteristic can be explained through both physical and mathematical perspectives [94]. Furthermore, low-cycle fatigue tests were conducted to assess the electrical and mechanical performance of the implant. During the fatigue study, the prototype underwent 30,000 axial loading cycles at a frequency of 1 Hz with a 350 N axial compression force. Fig. 2i shows the fatigue test results. The elastic modulus of the implant decreased by approximately 24% to 35.9 kPa after 30,000 loading cycles (see Fig. S1 in Supplementary Materials). The transmitted and received voltage values exhibited a decline from 1.1 V and 0.58 V in the initial 20,000 cycles to 0.87 V and 0.43 V, respectively. The transmitted and received voltage exhibited stability beyond 20,000 cycles. Nevertheless, the signal delivery ratio remained consistently around 0.5. This observed voltage trend reflects changes in both the mechanical and electrical properties of the wireless metamaterial implant. The diminishing electrical and mechanical performance under repeated loading cycles is anticipated and warrants careful investigation to establish calibration parameters for implants. However, the preferred performance for such implants varies case by case, and the target performance need not necessarily prioritize maximum electrical output or mechanical prowess, as it heavily relies on clinical requirements [7].

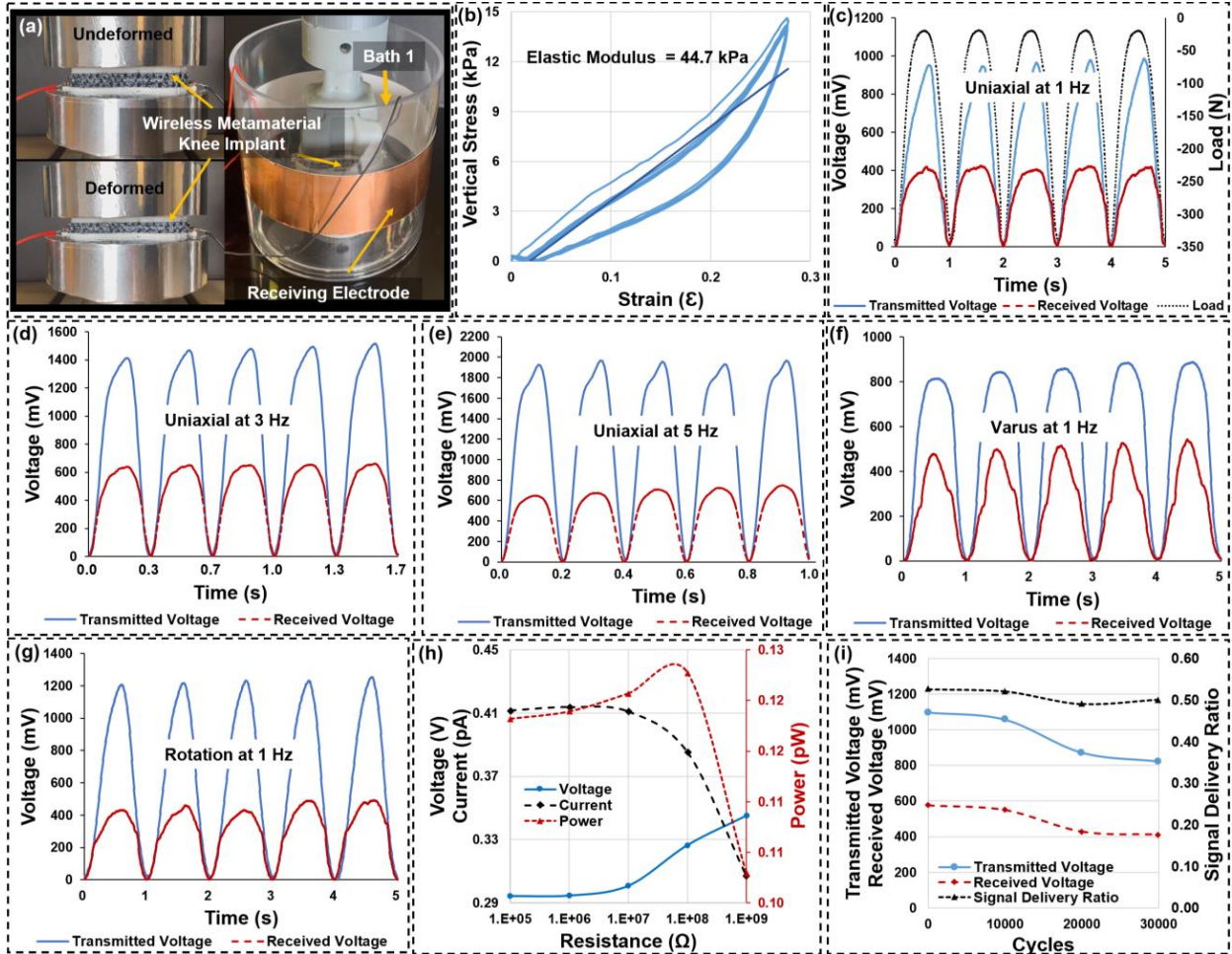

**Fig. 2. Experimental results showing:** (a) 3D printed wireless metamaterial TKR implant and test setup. (b) Stress–strain curves used to determine the elastic modulus of the implant. Electrical signals generated by the implant (in blue) and received by the electrode (in red) in air under uniaxial loading at (c) 1 Hz, (d), 3 Hz, (e) 5 Hz. Electrical signals generated by the implant (in blue) and received by the electrode (in red) in air under (f) varus loading, (g) 25° internal rotation. (h) Power-voltage-current curve output of the implant. (i) Low-cycle fatigue test results showing the variations of the signal delivery ratios (in black) over time.

Establishing the proposed wireless force sensing and communication framework requires theoretical characterizing the polarization, $P_s$, created by the electrostatic surface charges within the wireless metamaterial lattice in different media. To this aim, wireless transmission tests of the implant were conducted in air, simulated body fluid (SBF), and porcine, a theoretical model was subsequently developed. The TKR implant was a proof-of-concept prototype and was not optimized for ingress protection. Thus, it was encased within a smaller polyacrylonitrile bath with a diameter of 12.5 cm to prevent contact with SBF and tissue (Fig. 3a). The tests in air were replicated using the new setup. The space between the two testing baths was then filled with SBF and porcine tissue. Figs. 3b and c present the simplified schematic of the formulated wireless transmission in the experiments. In particular, the TKR implant and receiving electrode form the positive and negative electrodes of capacitor, while the dielectric contains four layers of mediums (Fig. 3c). The voltage and charge generated by the TKR implant can be determined using the *V-Q-x* relationship [6]. For the wireless metamaterial TKR implant with a built-in contact-separation mode TENG, the electric potential difference ($\Delta\varphi$) between the implant and receiving electrode can be expressed as:

$$\Delta\varphi = \int_{\rho_0}^{\rho_{\text{exter}}} \frac{\rho_0 V(t)}{\varepsilon_r^i \rho^2} d\rho$$

$$= \rho_0 V(t) \left[ \frac{1}{\varepsilon_r^1} \left( \frac{1}{\rho_0} - \frac{1}{\rho_{\text{inter}} - \gamma_{\text{acr}}} \right) + \frac{1}{\varepsilon_r^2} \left( \frac{1}{\rho_{\text{inter}} - \gamma_{\text{acr}}} - \frac{1}{\rho_{\text{inter}}} \right) \right. \tag{6}$$

$$\left. + \frac{1}{\varepsilon_r^3} \left( \frac{1}{\rho_{\text{inter}}} - \frac{1}{\rho_{\text{exter}} - \gamma_{\text{acr}}} \right) + \frac{1}{\varepsilon_r^4} \left( \frac{1}{\rho_{\text{exter}} - \gamma_{\text{acr}}} - \frac{1}{\rho_{\text{exter}}} \right) \right],$$

where $V$, $\varepsilon_r$, $\gamma_{\text{acr}}$, $\rho_{\text{inter}}$ and $\rho_{\text{exter}}$ are the output voltage, and relative dielectric constant, thickness of polyacrylonitrile baths, radius of internal polyacrylonitrile bath, radius of external polyacrylonitrile bath, and radius of TKR implant spherical shell, respectively. The details of the theoretical model derivation are presented in Materials and Methods. Table S1 in Supplementary Materials summarizes the geometric and material properties used in the theoretical modeling. Figs. 3d-f show the transmitted and received voltage values measured during the experiments and those predicted using the theoretical models for the air, tissue and SBF media, respectively. There is an acceptable agreement between the experimental and theoretical results. Fig. 3g shows the signal delivery ratios in air, tissue and SBF. The experimental delivery ratios in air, tissue and SBF are 0.55, 0.75 and 0.81, respectively. The delivery ratios estimated using the theoretical model are 0.54, 0.74, 0.74 in the respective media. An important observation from the results is that attenuation of the polarization electric field generated by the implant is lower in lossy (e.g. salt water, blood, animal tissue) than in lossless (e.g. air, vacuum) media. The ability of the polarization electric field to travel more effectively in lossy (conductive) media compared to lossless (non-conductive) media can be explained by the presence of free charge carriers in the conductive medium. In lossy media, there are mobile charged particles (ions) that can move in response to the applied electric field. These free charge carriers facilitate the transmission of the electric field through the medium, resulting in lower resistance and less dissipation of the signal. In contrast, lossless media like air or vacuum have fewer or no free charge carriers, leading to higher electrical resistance. As the polarization electric field travels through these non-conductive media, it encounters increased resistance, causing more rapid dissipation and attenuation of the signal. Following the procedure recommended in [95], signal-to-noise ratio (SNR) was measured to compare the strength of the signal wirelessly received by the electrode to the background noise for different media. The corresponding SNR of the received signal for air, porcine tissue, and SBF was 50.84 dB, 17.04 dB, and 16.69 dB, respectively. The increase in noise levels for each of the wirelessly received signals in lossy media can be attributed to their complex structure and ionic content.

In addition, we evaluate the wireless force sensing capability of the TKR implant in case of implant failure. To this aim, the intact implant was initially tested under uniaxial loading in air at 1 Hz. The transmitted and received signals for the intact implant are shown in Fig. 3h. Then, damage was introduced by cutting one of the unit cells of the implant through its width. The damaged implant was tested under uniaxial loading similar to the intact prototype. Fig. 3i presents the transmitted and received signals for the damaged implant. The transmitted and received signals sharply decrease from 0.77 V and 0.41 V to 0.3 V and 0.15 after introducing the damage, respectively. The signal delivery ratios are 0.53 and 0.51 for the intact and damaged implant, respectively. The decrease in the generated voltage is anticipated due to the presence of inactive unit cells that do not contribute to signal generation. Wireless self-sensing using the metamaterial systems has a broader range of applications for monitoring the health of structural systems experiencing multilevel damage states.

To further validate the functionality of the proposed TKR implant, a series of experiments were conducted using a human cadaver knee specimen. An *ex vivo* knee from a 75-year-old male donor (weight: 81.65 kg, height: 177.8 cm) was obtained with prior institutional ethics committee approval (Fig. 3j). Using CT scans of the specimen, the geometry of the TKR components were determined. These patient-specific components were then 3D printed. A series of experimental and FE simulations were carried out to assess the mechanical properties of the TKR implant. These measurements yielded the following values for the experimental compressive elastic modulus, torsional modulus and shear modulus of the TKR implant: 0.72 MPa, 1.28 MPa and 1.21 MPa, respectively. The implant's elastic modulus matched the target range for the human medial meniscus (20-80 kPa) [89,90]. Additionally, the implant has a Poisson's ratio of 0.1. Materials with a Poisson's ratio between roughly -0.1 and +0.1 are considered near-zero Poisson's ratio materials [96,97]. This high level of mechanical tunability and near-zero Poisson's ratio are unique characteristics of mechanical metamaterials, distinguishing them from conventional materials [98-101]. Details of the experimental and FE simulation results are presented in Supplementary Materials. The results demonstrate good agreement between the experiments and numerical simulations.

Standard surgical implantation techniques were used by an orthopedic surgeon to insert appropriately sized TKR components into the specimen (Fig. 3j). The specimen was then mounted in a custom-designed fixture to simulate knee flexion at 1 Hz with a maximum amplitude of 100 N (Fig. 3k). The electrode configurations are shown in Fig. 3k. To investigate the effect of electrode size on the received signals, four copper electrodes with varying surface areas (1 cm², 4 cm², 9 cm², and 16 cm²) were fabricated. These electrodes were then positioned at a fixed distance of 5 cm from the implant on the posterior aspect of the knee. This setup ensured all other parameters remained constant during the experiment. Video S5 in Supplementary Materials shows typical signals transmitted by the TKR implant within the cadaver model and received by the 9 cm² electrode. Fig. 3l shows the amplitude of the transmitted signals, the received signals, and the signal delivery ratio for the 1 cm², 4 cm², 9 cm², and 16 cm² electrodes. While the transmitted signal remains consistent around 0.35 V ± 10 mV for all electrodes under the same loading condition, the received signal strength significantly increases (205%) as the electrode surface area expands from 1 cm² to 16 cm². The received signal strength shows a minimal increase between electrodes of 1 cm² and 4 cm². However, it rises by 42% when comparing the 1 cm² and 9 cm² electrodes. In order to investigate the impact of electrode distance on the signal delivery ratio, the 9 cm² electrode (3 cm × 3 cm) was chosen as the reference electrode. This electrode was positioned at varying distances from the implant (0 cm (not attached to the skin), 2 cm, 5 cm, 10 cm, 20 cm). The results shown in Fig. 3l imply that for a constant loading amplitude, the signal delivery ratio exhibits a significant decrease as the electrode is moved away from the implant (adjacent to the skin) to a distance of 5 cm. Thereafter, the ratio stabilizes. Importantly, the received signal remains strong enough for detection at all distances tested. The SNR of the signals wirelessly received at distances of 0 cm, 2 cm, 5 cm, 10 cm, and 20 cm from the implant was 25.31 dB, 24.28 dB, 22.86 dB, 17.17 dB, and 16.69 dB, respectively.

A critical takeaway from the cadaveric study is the dependence of the received signal on both electrode size, distance and positioning. For consistent wireless force sensing measurements with this technology at the current stage, controlled conditions are necessary. This means using the same type of electrode and maintaining consistent electrode placement relative to the implant for each patient to ensure reliable comparisons with the "reference baseline" voltage signal. Consistent electrode placement refers to maintaining a fixed distance and location (anterior, posterior, or around the knee). The reference baseline voltage signal is the wirelessly received signal measured post-surgery. Subsequent readings for each prescribed motion pattern will be compared to this baseline. This essentially establishes a "relative" monitoring system where each subsequent signal is evaluated against the initial reference baseline. Furthermore, to achieve consistent measurements and facilitate meaningful comparisons, therapists can design a range of flexion and extension knee motions tailored to the patient's specific condition and treatment plan. This leverages the implant's ability, as shown in Fig. 2, to generate unique voltage outputs for different loading conditions. By comparing these standardized motions' voltage outputs to the reference baseline, deviations can reveal potential changes in knee biomechanics.

However, the presented proof-of-concept TKR implants demonstrate the first generation of implants with direct wireless communication capabilities based on the strain–induced Maxwell's displacement current. In this context, "direct" indicates that the same signal generated by the implant is directly used for wireless communication without relying on additional electronics, external power sources, or data loggers. Although the implant is not optimized for maximum power output, it enables mid-range wireless communication in real-time, even with low power output (< 1 pW). The wireless metamaterial implants do not resemble traditional electronic systems; they are biomaterial systems using their fabric for sensing, energy harvesting and wireless transmission. Furthermore, among the existing smart implants, the wireless metamaterial implants are the only class of mechanically tunable implants due to their metamaterial nature (see Fig. 1a). Mechanical tunability in orthopedic implants is crucial for customizing the implant to match individual patient biomechanics, ensuring optimal fit and function [7]. This adaptability also reduces the risk of complications and stress shielding and promotes long-term stability during the healing process [7]. The proposed wireless force sensing concept extends beyond orthopedic and spinal implants due to the scalability of the wireless metamaterial system. Eliminating the need for bulky wireless interrogation circuits can be a breakthrough in developing wireless robotic systems and medical implants. For instance, a wireless metamaterial cardiac stent featuring wireless force sensing capabilities can address a significant challenge associated with integrating electronics into the confined space within an artery. This innovation enables continuous monitoring of local hemodynamic changes, particularly during instances of restenosis.

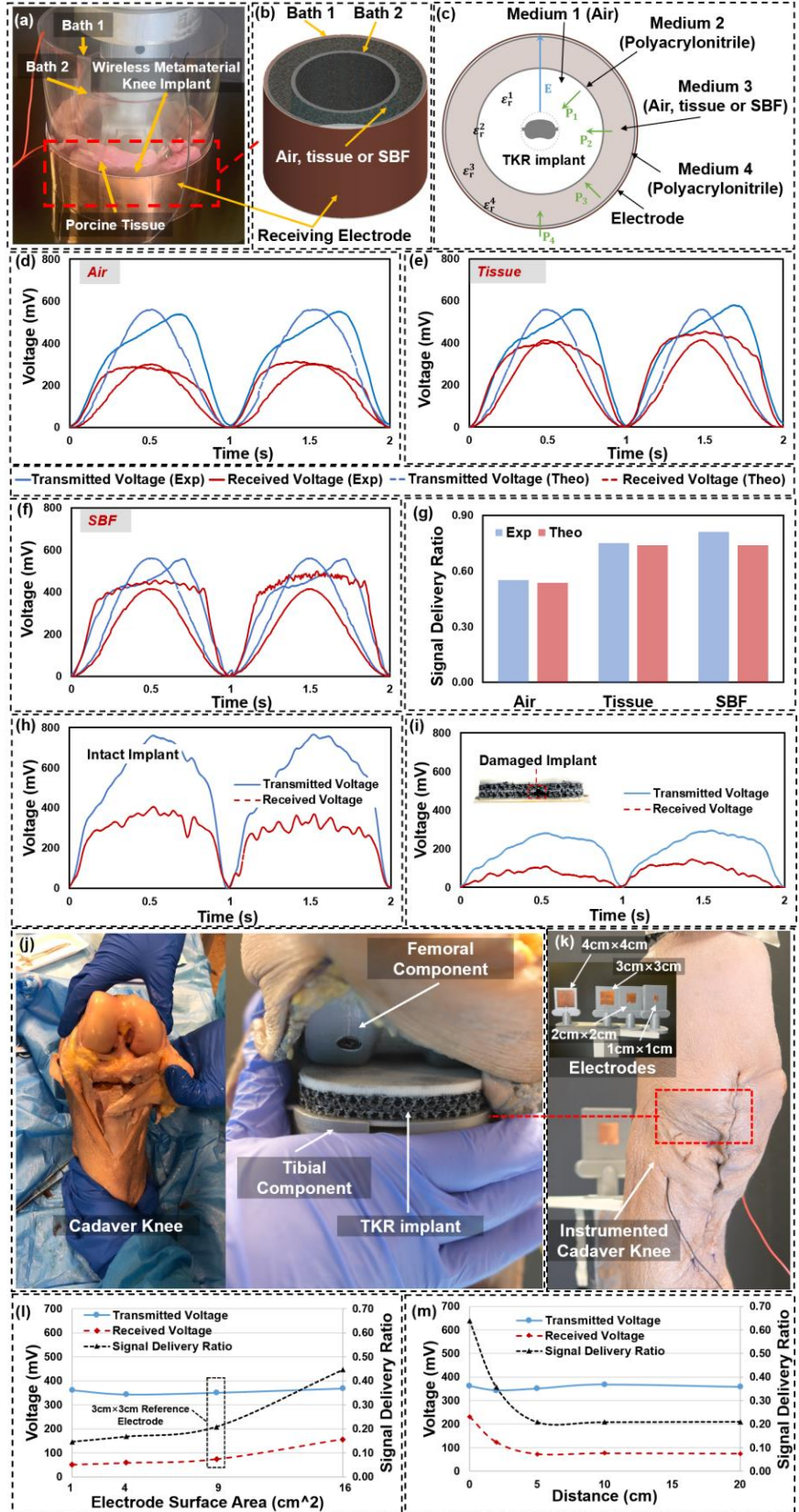

**Fig. 3. Experimental and theoretical evaluation of the wireless force sensing capability of the TKR implant.** (a) Experimental test setup using the porcine tissue. (b)(c) 3D and top view of the simplified schematic of the formulated wireless transmission in the experiments. Transmitted (in blue) and received (in red) electrical signals measured during the experiments and predicted using the theoretical model in (d) air, (e) porcine tissue, (f) SBF. (g) Signal delivery ratios in air, porcine tissue and SBF. (h) Transmitted (in blue) and received (in red) electrical signals generated by the intact implant in air under uniaxial loading. (i) Transmitted (in blue) and received (in red) electrical signals generated by the damaged implant in air under uniaxial loading. (j) Human cadaver knee before and after instrumentation with the 3D printed TKR components. (k) Experimental test setup using the cadaver knee and copper electrodes with varying surface areas. (l) Transmitted electrical signals (in blue), received electrical signals (in red) and signal delivery ratios (in black) for the 1 cm², 4 cm², 9 cm², and 16 cm² electrodes placed at a fixed distance of 5 cm from the implant on the back side of the cadaver knee. (j) Transmitted electrical signals (in blue), received electrical signals (in red) and signal delivery ratios (in black) for the reference 9 cm² electrode positioned at distances of 0 cm, 2 cm, 5 cm, 10 cm, and 20 cm from the implant on the back side of the cadaver knee.

**Conclusion**

We showed the feasibility of using wireless, electronic-free metamaterial implants for wireless transmission of the data in real-time. In the proposed alternative communication approach, the electrical signals harvested by the implants from mechanical stimuli are directly used for wireless transmission of the sensed data, eliminating the need for additional electronics, external power sources, or data loggers. A proof-of-concept TKR implant is created under the wireless metamaterial paradigm to evaluate the feasibility of the proposed wireless force sensing approach. The implant is tested under various loading conditions at 1, 3 and 5 Hz to establish its mechanical and electrical properties. The wireless transmission capability of the implant is then tested in air, SBF and porcine tissue. It is found that the wireless metamaterial implant with low power output on the order of 0.1 pW directly enables mid-range wireless communication without an external power source. An acceptable agreement is observed between the experimental and theoretical results. The results imply that the attenuation of the polarization electric field generated by the implant is lower in lossy than in lossless media. This can be regarded as an advantage of using the proposed mechanism *in vivo*. It is also shown that the wireless metamaterial implants can offer wireless self-sensing capability in case of implant failure. Experimental and numerical simulations were carried out to characterize the mechanical properties of the TKR implant. Further validation of the proposed technology was performed using a human cadaver knee specimen. Cadaver studies confirmed the dependence of the wireless signal strength on electrode size and placement. For reliable clinical use, consistent electrode selection and placement relative to the implant are crucial. This ensures accurate comparisons between the initial baseline signal and subsequent measurements during various motions. The system essentially functions by monitoring signal changes relative to this established baseline.

Further research should be conducted to optimize the electrical output of the implants for increasing the wireless transmission range. For instance, using biocompatible materials with high electrical conductivity, such as titanium (electrical conductivity ≈ 2.5 × 10⁶ S/m), to fabricate the conductive lattice of the implant could substantially increase the generated power. To precisely capture knee joint forces from different directions, the TKR implant can be segmented. Each segment could operate at a unique frequency or generate a distinct power output, allowing for the identification of the force source within the knee. A current limitation of these implants is their susceptibility to electromagnetic interference (EMI) in environments with strong EMI sources, such as MRI machines. This can lead to unreliable readings. Further research opportunities exist to explore mitigation strategies against EMI. Promising avenues include signal filtering techniques to isolate implant signals from strong EMI environments and active shielding using coils to generate magnetic fields that cancel out unwanted external fields. Future work can also focus on assessing different types of electrodes, including standard electrocardiogram (ECG) recorders and wearable recorders. Our future research will focus on evaluating the wireless transmission functionality of a series of optimized wireless metamaterial implants for *in vivo* testing in large animal models.

**Materials and Methods**
**Fabrication and testing of the TKR implants.** In this study, thermoplastic polyurethane (TPU) and polylactic acid (PLA) containing carbon black were employed to construct the dielectric and conductive lattices of the TKR implant, respectively. We exclusively used these biocompatible materials because they have been extensively studied and proven safe for implant applications [102]. To maintain their biocompatibility, the base materials were not modified. The elastic moduli and Poisson's ratios of TPU and PLA are 12 MPa, 3000 MPa, 0.48, and 0.25, respectively. PLA and TPU are positioned on the negative and positive ends of the triboelectric series, respectively, facilitating optimal electrification between the layers. The dielectric and conductive layers of the proposed implant were simultaneously produced in three distinct segments using the fused deposition modeling (FDM) technique and a Raise3D Pro2 Dual Extruder 3D Printer. The lattice structure of the metamaterial TKR implant was designed with parallel snapping segments with a semicircular shape to enable self-recovery behavior. The lattice is composed of multiple bi-stable unit cells, which include thick horizontal and vertical components along with a thinner curved section. The snapping segments were centrally anchored by more rigid supporting elements. The curved parts were mathematically designed using trigonometric functions to ensure a smooth snap-through transition and symmetrical stability. An NI9220 module with 1GΩ impedance and an SR570 amplifier were interfaced with the LabView software to measure the generated voltage and current values, respectively. The implants were mechanically tested using an Instron 8874 universal testing machine. The 3D modeling of the TKR implant designs was performed using AutoCAD and SolidWorks. All layers of the TKR implants (i.e. electrodes and dielectric layers) were printed simultaneously. Due to the noise from electrical interference (e.g., high-frequency electromagnetic waves), a low-pass filter was used to remove high-frequency noise from the original received signals. Origin 2018 was utilized to filter the data. In order to construct the power-voltage-current curve, the implant was connected to a circuit with adjustable resistors to systematically modulate the resistive load. The voltage and current produced by the implant were then measured at each resistance setting to create the power-voltage-current curve. By multiplying the recorded voltage and current values at each data point, power density was calculated. In the human cadaveric study, the orthopedic surgeon implanted the 3D-printed TKR components into the specimen. A midline incision was made over the knee. The soft tissue was dissected to the patella. The quadriceps tendon was transected proximal to the patella to expose the joint space. The anterior and posterior cruciate ligaments were resected while keeping the lateral and medial longitudinal ligaments intact. The proximal tibia and the menisci were resected. Then the distal femur was shaped. The implants were secured with screws and the incision was closed with sutures. The stands for the copper electrodes and the fixture holding them were printed using PLA. This cadaveric study was approved by the Committee for Oversight of Research and Clinical Training Involving Decedents (CORID) at the University of Pittsburgh, the ethical body responsible for reviewing research involving decedents and human cadaver specimens. CORID sanctioned the use of a cadaver knee for this research.

**Electrical characterization of the TKR implants.** The TKR implant features a built-in contact-separation mode TENG. The $V$-$Q$-$x$ relationship for this mode can be expressed as [103]:

$$V = -\frac{Q}{\varepsilon_0 S}\left[\frac{\gamma}{\varepsilon_r} + X(t)\right] + \frac{\sigma X(t)}{\varepsilon_0} \tag{7}$$

where, $V$, $Q$, $\varepsilon_0$, $\gamma$, and $\varepsilon_r$ are the output voltage, transferred charge, vacuum permittivity, thickness and relative dielectric constant, respectively. The varying gap distance $X(t)$ is given by the axial displacement $d(t)$. The surface charge density $\sigma$ is experimentally calibrated as 0.00136 μC/m². The contact area can be written as:

$$S = \sum_{i=1}^{n} \pi \rho_i W_i \tag{8}$$

where, $\rho_i$ and $W_i$ present the curvature radius of semicircular segment and width of dielectric layer of snapping units, respectively. Connecting a load $R$ to the TKR implant to form a circuit, the output voltage can be calculated by Ohm's law as:

$$V = R\frac{dQ}{dt} \tag{9}$$

Substituting Eq. (9) into Eq. (7), we have:

$$\frac{dQ}{dt} = -\lambda Q + \frac{\sigma X(t)}{\varepsilon_0 R} \tag{10}$$

where,

$$\lambda = \frac{\gamma + \varepsilon_r X(t)}{\varepsilon_0 \varepsilon_r RS} \tag{11}$$

Solving the differential equation Eq. (10) yields:

$$Q(t) = e^{-\int \lambda \, dt} \int \frac{\sigma X(t)}{\varepsilon_0 R} e^{\int \lambda \, dt} \, dt + \mu e^{-\int \lambda \, dt}, \tag{12}$$

where, $\mu$ denotes the integration constant. Substituting Eq. (12) into Eq. (9) leads to:

$$V(t) = -\lambda e^{-\int_0^t \lambda \, d\xi} \int_0^t \frac{\sigma X(\xi)}{\varepsilon_0} e^{\int_0^\xi \lambda \, d\tau} d\xi + \frac{\sigma X(t)}{\varepsilon_0} - \mu R \lambda e^{-\int_0^t \lambda \, d\xi}. \tag{13}$$

The boundary conditions of Eq. (10) are given as:

$$\begin{cases} Q|_{t=0} = 0 \\ V|_{t=0} = -\frac{Q|_{t=0}}{\varepsilon_0 S}\left[\frac{\gamma}{\varepsilon_r} + X|_{t=0}\right] + \frac{\sigma X|_{t=0}}{\varepsilon_0} = \frac{\sigma X|_{t=0}}{\varepsilon_0} - \mu \frac{\gamma + \varepsilon_r X|_{t=0}}{\varepsilon_0 \varepsilon_r S} \end{cases} \tag{14}$$

$\mu$ is determined as 0 according to Eq. (14), and thus, $Q(t)$ and $V(t)$ can be rewritten as:

$$Q(t) = e^{-\int \lambda \, dt} \int \frac{\sigma X(t)}{\varepsilon_0 R} e^{\int \lambda \, dt} \, dt \tag{15}$$

and

$$V(t) = -\lambda e^{-\int_0^t \lambda \, d\xi} \int_0^t \frac{\sigma X(\xi)}{\varepsilon_0} e^{\int_0^\xi \lambda \, d\tau} d\xi + \frac{\sigma X(t)}{\varepsilon_0} \tag{16}$$

For the open-circuit condition, the load $R$ can be treated as infinity, Therefore, $Q(t)$ and $V(t)$ are:

$$Q(t) = 0, \tag{17}$$

and

$$V(t) = \frac{\sigma X(t)}{\varepsilon_0}. \tag{18}$$

Based on the principle of signal attenuation in the wireless transmission process (as illustrated in Fig. 3c), Eq. (4) can be formulated as follows:

$$\boldsymbol{D} = \varepsilon_0 \boldsymbol{E} + \sum_{i=1}^{4} \boldsymbol{P}_i \tag{19}$$

where, $\boldsymbol{P}_i$ ($i = 1, 2, 3, 4$) presents the polarization electric field in medium $i$ with the presence of electric field $\boldsymbol{E}$. The Maxwell's displacement current density can be written as:

$$\boldsymbol{J}_D = \varepsilon_0 \frac{\partial \boldsymbol{E}}{\partial t} + \sum_{i=1}^{4} \frac{\partial \boldsymbol{P}_i}{\partial t} \tag{20}$$

In order to quantitatively analyze the signal attenuation in the wireless transmission process, the external electric field of TKR implant is simplified as the electric field of a uniformly charged spherical shell with the charge of $q$ and radius of $\rho_0$. $\rho_0$ is experimentally calibrated as 35 mm. In the vacuum, the electric field strength outside the spherical shell is:

$$\boldsymbol{E}(\rho) = \frac{q}{4\pi \varepsilon_0 \rho^2} \tag{21}$$

where, $\rho$ ($\rho > \rho_0$) denotes the distance from the center of spherical shell. The transmitted voltage $V(t)$ can be regarded as the surface electric potential of spherical shell, and thus, we have:

$$V(t) = \frac{q}{4\pi \varepsilon_0 \rho_0} \tag{22}$$

substituting Eq. (22) into Eq. (21), $E(\rho)$ is rewritten as:

$$\boldsymbol{E}(\rho) = \frac{\rho_0 V(t)}{\rho^2} \tag{23}$$

In order to find the combined electric field of $\boldsymbol{E}$ and $\boldsymbol{P}_i$ ($i = 1, 2, 3, 4$), Eq. (1) can be expressed as:

$$\boldsymbol{E}' = \frac{\boldsymbol{E}}{\varepsilon_r^i} \tag{24}$$

where $\varepsilon_r^i$ ($i = 1, 2, 3, 4$) is the relative permittivity of medium $i$. Thence, the actual electric field strength is modified as:

$$\boldsymbol{E}'(\rho) = \frac{\rho_0 V(t)}{\varepsilon_r^i \rho^2} \tag{25}$$

The electric potential difference, $\Delta \varphi$, is defined as the integral of electric field strength over the effective displacement, as expressed in Eq. (6).

**Supplementary Materials**
Supporting Information is available from the online library or from the author.

**Video S1.** Video showing deformation of the TKR implant under uniaxial loading at 1 Hz
**Video S2.** Video showing electrical signals generated by the implant and received by the electrode under uniaxial loading at 1 Hz
**Video S3.** Video showing electrical signals generated by the implant and received by the electrode under uniaxial loading at 3 Hz

**Video S4.** Video showing electrical signals generated by the implant and received by the electrode under uniaxial loading at 5 Hz
**Video S5.** Video showing electrical signals generated by the implant within the human cadaver model and received by the electrode under uniaxial loading at 1 Hz.

**Acknowledgments**
Research reported in this work was supported by the National Science Foundation (NSF) CAREER award (CMMI-2235494) and the National Institute of Biomedical Imaging and Bioengineering (NIBIB) of the National Institutes of Health (NIH) under award number 1R21EB034457-01A1. This research is a continuation of U.S. Provisional Pat. Ser. No. 63/330,156 and No. 63/420,309 filed in 2022 by A.H.A..

**Author Contributions**
A.H.A.: conceived the concept. J.L., W.L., D.J., and W.M.: performed the experiments. J.L., K.B., J.W., and P.J. performed numerical and theoretical studies. J.L., W.L. and W.M.: carried out the fabrications supervised by A.H.A. and K.B.: R.P.K. and N.A.: Contributed to the design of the clinically relevant TKR implant and experiments, and interpretation of the data. All authors analyzed and interpreted the data. The manuscript was written by J.L., D.J., A.H.A. and N.A.: All authors contributed to editing the manuscript.

**Conflicts of Interest**
There are no conflicts of interest to declare.

**Data Availability**
The data that support the findings of this study are available from the corresponding author upon reasonable request.

# Supplementary Materials for

# Wireless Electronic-free Mechanical Metamaterial Implants

Jianzhe Luo[1,2], Wenyun Lu[1,2], Pengcheng Jiao[3], Daeik Jang[2], Kaveh Barri[4], Jiajun Wang[3], Wenxuan Meng[2], Rohit Prem Kumar[5], Nitin Agarwal[5,6,7], D. Kojo Hamilton[5,6,7], Zhong Lin Wang[8,9,10], Amir H. Alavi[1,2,a]

[1]*Department of Bioengineering, University of Pittsburgh, Pittsburgh, PA, USA*

[2]*Department of Civil and Environmental Engineering, University of Pittsburgh, Pittsburgh, PA, USA*

[3]*Ocean College, Zhejiang University, Zhoushan, Zhejiang, China*

[4]*Department of Civil and Systems Engineering, Johns Hopkins University, Baltimore, MD, USA*

[5]*Department of Neurological Surgery, University of Pittsburgh School of Medicine, Pittsburgh, PA, USA*

[6]*Department of Neurological Surgery, University of Pittsburgh Medical Center, Pittsburgh, PA, USA*

[7]*Neurological Surgery, Veterans Affairs Pittsburgh Healthcare System, Pittsburgh, PA, USA*

[8]*Beijing Institute of Nanoenergy and Nanosystems, Chinese Academy of Sciences, Beijing, China*

[9]*School of Materials Science and Engineering, Georgia Institute of Technology, Atlanta, GA, USA*

[10]*Yonsei Frontier Lab, Yonsei University, Seoul 03722, Republic of Korea*

---

[a]Email: alavi@pitt.edu (A.H. Alavi)

## S1. Mechanical characterization results

Figs. S1a-c show the elastic modulus of the TKR implant after 10,000, 20,000 and 30,000 loading cycles. Figs. S1d shows the transverse versus longitudinal strains used to determine the Poisson's ratio of the TKR implant. In order to track the strain fields in the loading and transverse directions, a video gauge Ncorr [1] was used. A camera (Sony Model ILCE-7M3) was used to record the video. Following the procedure explained in [2], the Poisson's ratio was calculated as 0.1.

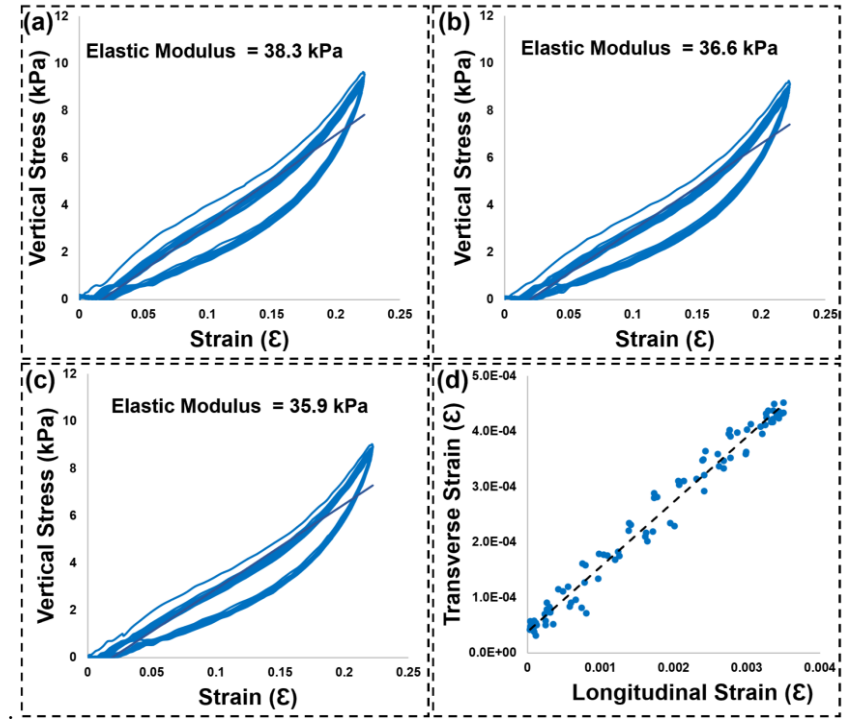

**Fig. S1.** Stress–strain curves used to determine the elastic modulus of the TKR implant after (a) 10000, (b) 20000, and (c) 30000 cycles. (d) Transverse versus longitudinal strains used to determine the Poisson's ratio of the TKR implant.

The mechanical behavior of the TKR implant used in the cadaveric study was experimentally tested and then numerically simulated in Abaqus/CAE using the static/general solving algorithm. Fig. S2a presents the meshing, boundary and loading conditions. The model was meshed into 273307 elements with the Tet shape and Free technique. The bottom parts were continuously clamped in numerical simulations. For the compression case, the upper parts were subjected to an axial compressive displacement of 0.6 mm (Fig. S2b). For the torsion case, the upper parts were subjected to a counterclockwise rotation of 2 degrees (Fig. S2c). For the shear case, the upper parts were subjected to a lateral displacement of 0.5 mm in the y direction (Fig. S2d). Fig. S2e compares the force-displacement and stress-strain relationships under compression between experimental and numerical results. Fig. S2f compares the torque-rotation and stress-strain relationships under torsion between experimental and numerical results. Fig. S2g compares the force-displacement and stress-strain relationships under shear between experimental and numerical results. The results reveal that the experiments and numerical simulations are in satisfactory agreement. According to the stress-strain relationships, the experimental compressive elastic modulus, torsional modulus and shear modulus of TKR implant are 0.72 MPa, 1.28 MPa, 1.21 MPa, respectively. The corresponding values from the FE simulations are 0.75 MPa, 1.34 MPa and 1.31 MPa, respectively.

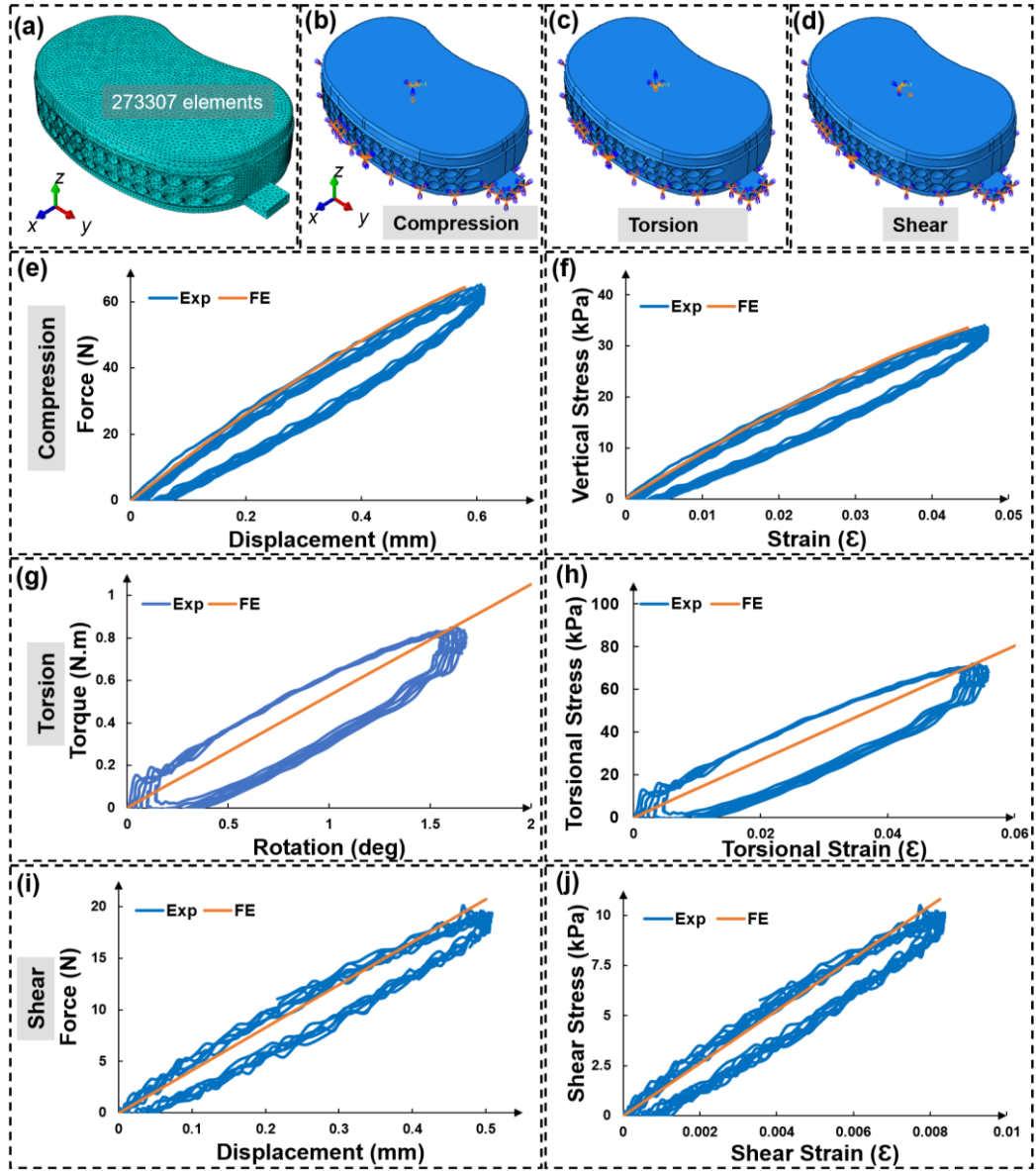

**Fig. S2: Mechanical characterization of the TKR implant.** (a) Meshing condition in FE simulations. Boundary and loading conditions for the (b) compression, (c) torsion and (d) shear cases in numerical simulations. Comparisons of (e) force-displacement and (f) stress-strain relationships under compression between experimental and numerical results. Comparisons of (g) torque-rotation and (h) stress-strain relationships under torsion between experimental and numerical results. Comparisons of (i) force-displacement and (j) stress-strain relationships under shear between experimental and numerical results.

## S2. Geometric and material properties

Table S1 summarizes the geometric and material properties used in the theoretical modeling.

**Table 1.** Geometric and material properties.

| | | |
|---|---|---|
| Geometric properties | Radius of simplified spherical shell $\rho_0$ (mm) | 35 |
| | Thickness of polyacrylonitrile baths $\gamma_{acr}$ (mm) | 4 |
| | Radius of internal polyacrylonitrile bath $\rho_{inter}$ (mm) | 50 |
| | Radius of external polyacrylonitrile bath $\rho_{exter}$ (mm) | 75 |
| Material properties | Relative permittivity of air $\varepsilon_r^{air}$ | 1.00053 |
| | Relative permittivity of polyacrylonitrile $\varepsilon_r^{poly}$ | 4.85 |
| | Relative permittivity of tissue $\varepsilon_r^{tis}$ | 48.5 |
| | Relative permittivity of SBF $\varepsilon_r^{SBF}$ | 55 |